\documentclass[conference]{IEEEtran}
\IEEEoverridecommandlockouts
\usepackage{xcolor}
\usepackage{mathtools}
\usepackage{epsfig,makeidx,color}
\usepackage{amsmath,amssymb,bbm,enumitem}
\usepackage{cite,graphicx,lipsum}
\usepackage[switch,pagewise]{lineno}
\usepackage{tabularx}
\usepackage{algorithm}
\usepackage{algpseudocode}
\usepackage{hyperref}
\usepackage{cases}
\usepackage{graphicx}
\usepackage{booktabs}
\usepackage{subcaption}
\usepackage{cleveref}

\newcommand*{\affaddr}[1]{#1} 
\usepackage{subcaption}
\newcommand*{\affmark}[1][*]{\textsuperscript{#1}}
\hypersetup{
        colorlinks = true,
        citecolor = red,
        breaklinks = true,
        linktocpage=true,
}
\usepackage{url}

\def\cM{{\cal M}}

\def\cK{{\cal K}}
\def\cX{{\cal X}}

\def\cG{{\cal G}}

\def\rT{{\rm T}}

\def\uR{{\mathbb R}}

\def\uN{{\mathbb N}}

\def\be{ \begin{equation} }
\def\ee{ \end{equation} }
\def\bea{ \begin{eqnarray} }
\def\eea{ \end{eqnarray} }

\def\bx{{\bf x}}

\def\bg{{\bf g}}

\def\bu{{\bf u}}

\def\bz{{\bf z}}

\def\bw{{\bf w}}

\def\bA{{\bf A}}

\def\bF{{\bf F}}

\def\bK{{\bf K}}

\def\b0{{\bf 0}}

\def\cN{{\cal N}}

\ifCLASSOPTIONonecolumn
\else

\fi

\usepackage{geometry}
\begin{document}

\title{Graph Koopman Autoencoder for Predictive Covert Communication Against UAV Surveillance
}


\author{%
Sivaram Krishnan\affmark[1], Jihong Park\affmark[1], Gregory Sherman\affmark[2], Benjamin Campbell\affmark[2], and  Jinho Choi\affmark[1] \\
\affaddr{\affmark[1]School of Information Technology, Deakin University,  Australia }\\
\affaddr{\affmark[2]Defense Science and Technology Group, Australia}\\

}

\maketitle
\begin{abstract}
Low Probability of Detection (LPD) communication aims to obscure the very presence of radio frequency (RF) signals, going beyond just hiding the content of the communication. However, the use of Unmanned Aerial Vehicles (UAVs) introduces a challenge, as UAVs can detect RF signals from the ground by hovering over specific areas of interest. With the growing utilization of UAVs in modern surveillance, there is a crucial need for a thorough understanding of their unknown nonlinear dynamic trajectories to effectively implement LPD communication. Unfortunately, this critical information is often not readily available, posing a significant hurdle in LPD communication.
To address this issue, we consider a case-study for enabling terrestrial LPD communication in the presence of multiple UAVs that are engaged in surveillance. We introduce a novel framework that combines graph neural networks (GNN) with Koopman theory to predict the trajectories of  multiple fixed-wing UAVs over an extended prediction horizon. Using the predicted UAV locations, we enable LPD communication in a terrestrial ad-hoc network by controlling nodes' transmit powers to keep the received power at UAVs' predicted locations minimized. Our extensive simulations validate the efficacy of the proposed framework in accurately predicting the trajectories of multiple UAVs, thereby effectively establishing LPD communication. 
\end{abstract}


\ifCLASSOPTIONonecolumn
\baselineskip 28pt
\fi

\section{Introduction}
In the evergrowing landscape of mobile surveillance and digital warfare, enabling covert communication is paramount \cite{wang2019secrecy}. While existing cryptography and physical layer security (PLS) focus on concealing the content of communication \cite{shiu2011physical}, covert communication aims to hide the existence of communication links from untrustworthy entities or adversaries. In wireless systems, this covert communication challenge is formulated as a problem of low probability of detection (LPD) \cite{yan2019low}, which seeks to minimize transmit power to evade signal detection while maintaining minimal connectivity.





\begin{figure}
    \centering
\includegraphics[width=.9 \columnwidth]{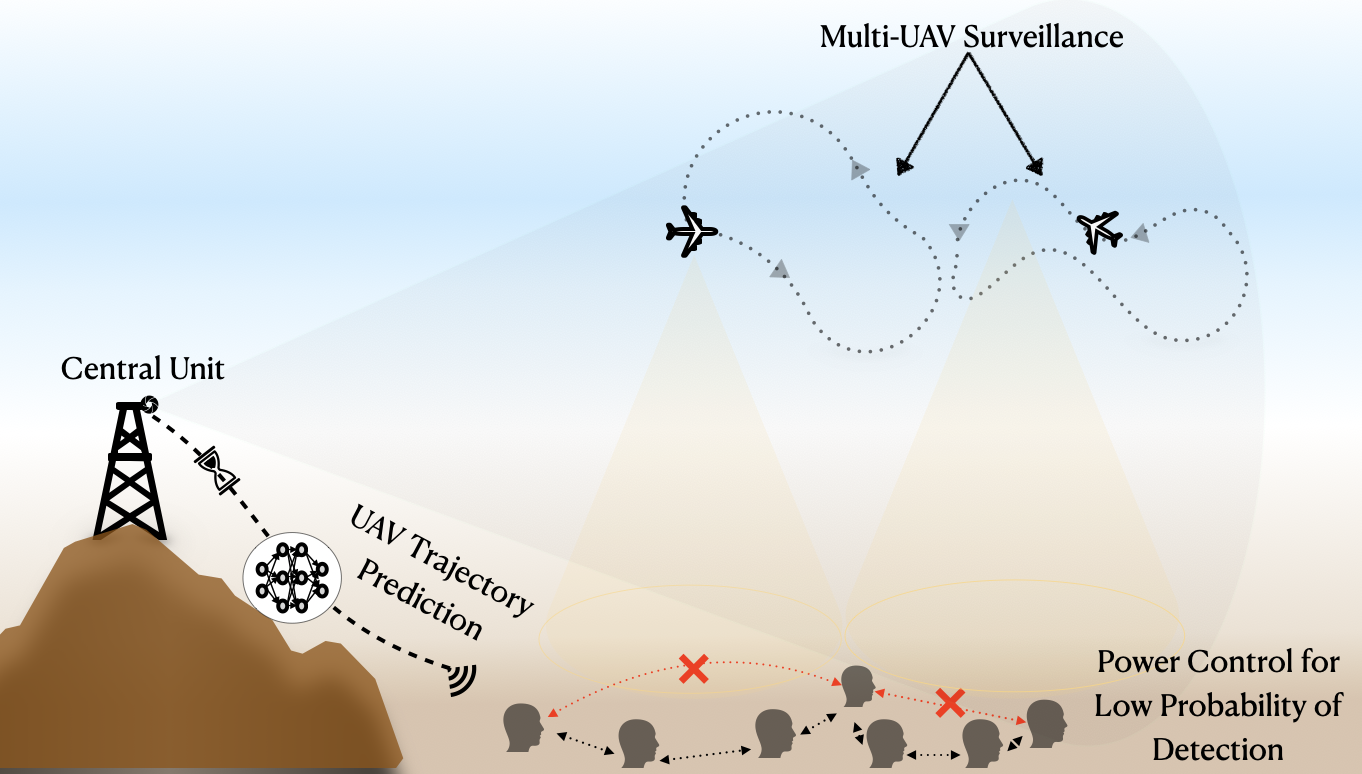}
    \caption{
    An illustration of covert communication against UAV surveillance (top), wherein ground nodes (bottom) adjust transmit power for achieving low probability of detection based on UAV trajectory prediction by their associated central unit (left).} 
    \label{fig:problem}
\end{figure}

The LPD problem has been extensively studied in point-to-point communication scenarios \cite{sobers2017covert, bash2013limits}, wherein Alice communicates with Bob under the surveillance of Willie. Extending this to ad-hoc networks, LPD with multi-hop connectivity constraints has been recast by the minimum-area spanning tree (MAST) problem \cite{carmi2006minimum, campbell2018minimising}. This approach aims to minimize the network's coverage area to minimize the probability of signal detection by unknown eavesdroppers. Several algorithms have been developed to solve the MAST problem, including branch-and-cut algorithms \cite{guimaraes2021minimum}, but their complexity increases with the number of nodes. To address this scalability issue, deep learning has been employed to solve the MAST problem with low latency, regardless of node quantity \cite{Krishnan_SSP23}. However, these existing works do not account for emerging mobile surveillance technologies, such as unmanned aerial vehicle (UAV) assisted surveillance \cite{wang2019secrecy, chen2021uav}, nor do they consider intelligent nodes capable of predicting surveillance movements.

To address these gaps, in this article we introduce a novel LPD scenario where a ground ad-hoc network aims to achieve LPD against a surveillance network of mobile UAVs, as illustrated in Fig.~\ref{fig:problem}. The ground users are coordinated by a central unit (CU) that predicts UAV mobility and conveys this information to ground users to optimize transmit power. This problem is non-trivial, especially considering the nonlinear and complex mobility patterns of UAVs \cite{shao2023path}. 

To cope with this new LPD challenge, we first develop a novel multi-UAV nonlinear dynamics prediction algorithm based on graph neural networks (GNNs) \cite{kipf2016semi, hamilton2017inductive} and Koopman operator theory \cite{Koopman31}, coined graph-based Koopman autoencoder (GKAE). In GKAE, the GNN spatially reduces a snapshot of multi-UAV network topology into a single latent state, while the Koopman operator based autoencoder linearizes and learns the temporal mobility pattern of these latent states, enabling long-term and scalable multi-UAV movement prediction with low latency. With these predicted UAV locations, we optimize transmit power using a branch-and-cut algorithm. Our proposed GKAE successfully predicts the trajectories of 4 UAVs over 80 future time slots with a mean-squared error of $0.0025$. This enables achieving LPD with the UAV's received power being at least 18\% lower than their signal detection threshold.

\section{Background: Koopman Operator Theory}\label{Sec:Koop}

Mobility prediction is severely challenged under nonlinear dynamics. Traditional linearization techniques such as Jacobian approximation is accurate within a short-term interval, hindering long-term prediction. Koopman operator theory \cite{Koopman31} offers an effective alternative, which linearizes not a single point but the subspace of entire points. To be specific, consider a state vector $\bx(t)\in \cX \subseteq \uR^N $ of a node at a discrete time $t \in \uN$. The nonlinear temporal dynamics of $\bx(t)$ can be described as
\be 
\bx(t+1) = \bF(\bx(t)),
    \label{EQ2:xFx}
\ee 
where $\bF$ is a flow map.
Consider that $g: \cX \to \uR$ is a measurement function. Koopman operator theory states that there exists a linear (infinite-dimensional) operator $\cK$ that acts to advance all $g$, i.e.,
\begin{align}
\cK g = g \circ \bF,
    \label{EQ3:KO}
\end{align}
where $\circ$ represents the composition operator and $\cM$ denotes a smooth manifold. 
Applying \eqref{EQ3:KO} to \eqref{EQ2:xFx}, we have
\begin{align}
g(\bx(t+1)) = g \circ \bF(\bx(t)) = \cK g(\bx(t)),
    \label{EQ:gg}
\end{align}
where $g(\bx(t))$ is an observable measured at time $t$.
This can be extended to the case with multiple observables. Precisely, let $\bg (t) = [g_1 (t) \ \ldots \ g_M (t)]^\rT$, where $g_m (t) = g_m (\bx(t))$. Then, from \eqref{EQ:gg}, we have
\be 
\bg(t+1) = \cK \bg(t). 
\ee 

If $\bg(t) \in \cG$ and $\cK \bg(t) \in \cG$, where $\cG$ is a finite-dimensional space, $\cG$ becomes a Koopman invariant subspace. In this case, $\cK$ becomes a finite-dimensional linear operator and is represented by a matrix $\bK \in \uR^{M \times M}$, which is called the Koopman matrix of dimension $M$. The eigenvalues and eigenvectors of $\bK$ describe the linear evolution of the dynamical system in the Koopman invariant subspace. A prerequisite to discovering $\bK$ is to find a Koopman invariant subspace. While this has been tackled traditionally using predefined kernel functions, recent frameworks rely on training an encoder-decoder structured deep neural network (DNN) \cite{lusch2018deep, Brunton22}, hereafter referred to as Koopman autoencoder (KAE).

\section{System Model and Problem Formulation}\label{sec:Sys}

In this section, we present our system model for achieving LPD communication against UAV surveillance. 

\subsection{Terrestrial Ad-hoc Network}

We consider a terrestrial wireless ad-hoc network with $N$ ground nodes, denoted by the set $\mathcal{N} = \{1, 2, \cdots, N\}$ at  known locations given by $\{{\bw_n : (w_{n_x}, w_{n_y}, 0)}\}_{n \in \mathcal{N}}$. Here, $\bw_n \in \mathbb{R}^{3}$ represents the three-dimensional (3D) coordinates of the $n$-th ground node. Each ground node operates with an adjustable transmit power, which is represented by $P_n(t)$ at time $t$, $\forall n \in \mathcal{N}$. Then, the corresponding signal-to-noise ratio (SNR) at receiver $j$, denoted by $\gamma_{i_{j}}$, becomes
\be
\gamma_{i_{j}}(t) = \frac{P_i(t)d_{i, j}^{-\eta} \nu_{i,j} (t)}{N_0}, \forall i \ne j \in \mathcal{N} ,
\ee
where $d_{i, j} = ||\bw_i - \bw_j||$ is the Euclidean norm denoting the distance between the $i$-th and $j$-th nodes, $\nu_{i,j} (t)$ is the small-scale fading term, and $\eta$ and $N_0$ represent the path-loss exponent and noise variance, respectively. 

To ensure a stable\footnote{Due to the presence of the small-scale fading term, $\nu_{i,j}(t)$, in terrestrial communication, the SNR becomes time-variant. Consequently, when the SNR undergoes deep fading, leading to low signal strength, an outage event may occur. Thus, to maintain stable communication, it is crucial to keep the outage probability low, even at the expense of reducing the data-rate transmission.} 
communication link between the ground nodes, the received SNR must exceed a predefined SNR threshold, represented by $\tilde{\gamma}$. Consequently, the set of communication links for node $i$ at time $t$ can be formally expressed as:
\be
C_i(t) = \{j : \gamma_{i_{j}}(t) \ge \tilde{\gamma}\}_{i \ne j \in \mathcal{N}}.
\ee

\subsection{UAV Surveillance Network}

We also consider the presence of $L$ UAVs, represented by the set $\mathcal{L} = \{1, 2, \cdots, L\}$ with time-variant locations $\{{\bu_l(t): (u_{l_x}(t), u_{l_y}(t), h)}\}_{l \in \mathcal{L}}$. These locations indicate the 3D coordinates of the $l$-th UAV, which hovers at a constant altitude denoted by $h$.

An air-ground channel is assumed to be dominated by line-of-sight (LoS). The received signal strength at the $l$-th UAV from ground node $n$ at time $t$ is given as
\be
P_{l, n}(t) = P_n(t) d_{l, n}(t)^{-\eta^\prime},
\ee
where $\eta^\prime$ is the path-loss exponent for air-ground channel.  

The presence of a group of multiple UAVs in a given area leads to spatial dependencies, with UAVs situated near each other actively communicating. This inter-UAV communication plays a crucial role in determining the dynamics of these autonomous UAVs.  The neighboring UAVs for UAV $k$, are defined by the set $\alpha_k(t)$ which is represented as
\be
\alpha_k(t) = \{l:||\bu_k(t) - \bu_l(t)|| \le \tilde{D}\}_{k \ne l \in \mathcal{L}},\label{eq:Nei}
\ee
where $\tilde{D}$ is the minimum distance threshold. Assuming the UAVs are of the fixed-wing type, the dynamics of UAV $l$ can be described as follows \cite{muslimov2021consensus}:
\begin{align}
\frac{u_{l_x}(t) - u_{l_x}(t-1)}{\Delta t} & = v_l \cos(\phi_l(t-1)) + v_{w} \cos(\theta_{w}) \label{EQ:ukx} \\
\frac{u_{l_y}(t) - u_{l_y}(t-1)}{\Delta t} & = v_l \sin(\phi_l(t-1)) + v_{w} \sin(\theta_{w})  \label{EQ:uky} \\
\phi_l(t) & = \phi_l(t-1) + 0.1\phi_{\text{agg}, l}(t-1) 
 \label{subeq:6}
\end{align}
where $v_l$ is the forward velocity, $v_w$ is the wind velocity while $\phi_l$ is the turning angle, and $\theta_w$ is the wind direction. 
Furthermore, the adjustment of the turning angle for the UAV, is influenced by its own rotational change, and the aggregated rotations of neighboring UAVs, $\forall k \in \alpha_l(t)$. The aggregated turning angle for UAV $l$ at time $t$ becomes
\be
\phi_{\text{agg}, l}(t) =\frac{1}{|\alpha_l(t)|}\sum_{k \in \alpha_l(t)} \phi_k(t) .
\ee


\begin{figure}
    \centering
\includegraphics[width=\columnwidth]{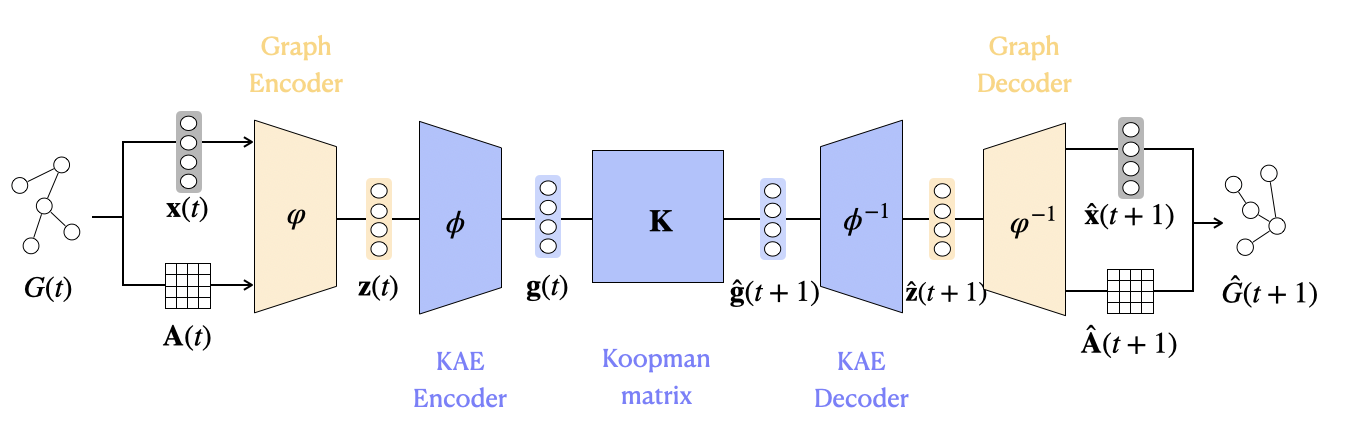}
    \caption{Graph-based Koopman autoencoder (GKAE) architecture.}
    \label{fig:Proposed}
\end{figure}

\section{Mobility Prediction and Power Control for LPD}
In this section, we present our optimization methodologies for addressing the problems $\textbf{P1}$ and $\textbf{P2}$, which focus on multi-UAV trajectory prediction and LPD optimization, respectively.

\subsection{GNN-Aided Koopman Prediction of Multi-UAV Trajectories}

We express the UAV network as a graph. Formally, a graph at time $t$ is defined as $G(t) = (\mathcal{V}, \mathcal{E}(t))$, where $\mathcal{V} = \{1, \cdots, L\}$ is the set of the nodes representing UAVs, $\mathcal{E}(t)$ is the set of edges encapsulating spatial dependencies among the UAVs at time $t$ which is determined based on \eqref{eq:Nei}. Additionally, let $\bu_{k}$ be the node features for node $k \in \mathcal{V}$, corresponding to the 3D coordinates of the $k$-th UAV. 


As illustrated in Fig.~\ref{fig:Proposed}, the GKAE is divided into two major components: GNN-based graph encoder-decoder and KAE encoder-decoder. First, the graph encoder-decoder architecture follows from the sample and aggregate graph convolution network (SAGE), which serves as an extension to the standard graph convolution network (GCN) \cite{kipf2016semi, hamilton2017inductive}. At the $\ell$th layer with weights $\boldsymbol{w}^{(\ell)}$, for the $k$-th node with a set $\alpha_k$ of neighbours, a SAGE layer yields the output activation $\boldsymbol{z}_k^{(\ell+1)}(t)$ as follows:
\begin{align}  
\label{eq:SAGEConv} \boldsymbol{z}_k^{(\ell+1)}(t) &= \sigma \left( \boldsymbol{w}^{(\ell)} \cdot \left[ \boldsymbol{z}_k^{(\ell)}(t) \Vert \bigoplus_{l \in \alpha_k(t)}  \left(\boldsymbol{z}_k^{(\ell)}(t), \boldsymbol{z}_l^{(\ell)}(t) \right) \right] \right) 
\end{align}
where $\sigma(\cdot)$ is the nonlinear activation function, $\bigoplus$ is an permutation invariant aggregation function which could be mean or max pooling and $\Vert$ is the concatenation operation. Using \eqref{eq:SAGEConv}, the graph encoder $\varphi$ reduces $G(t)$ to a {single} latent variable, denoted as $\bz(t) = [\boldsymbol{z}_1 (t)^\top \cdots \boldsymbol{z}_L(t)^\top]^\top$. 
Next, by receiving $\bz(t)$ as its input, the KAE encoder $\phi$ produces $\bg(t)$, and the KAE decoder $\phi^{-1}$ yields an one-step time lagged output $\bz(t+1)$. A Koopman matrix $\bK$ connects $\phi$ and $\phi^{-1}$, performing a one-step time lag $\bg(t+1) = \bK \bg(t)$ in the Koopman invariant subspace.


To train GKAE, we define loss function $L_{GKAE}$ as the following weighted sum of two loss terms:
\begin{align}
\textbf{(P1) :} \quad \min_{\boldsymbol{\varphi}, \mathbf{K}, \boldsymbol{\psi}, \boldsymbol{\varphi}^{-1}, \boldsymbol{\psi}^{-1}} \beta_1 L_{\text{rec}} + \beta_2 L_{\text{pred}},
    \label{eq:minloss}
\end{align}
where $\beta_1$ and $\beta_2$ are hyperparameters.
Here,
\begin{align}
L_{\text{rec}} & = \sum_{t=1}^T ||\tilde \bx(t)||^2 + ||\tilde \bz(t) ||^2 \label{eq:lossre} \\
L_{\text{pred}} & = \sum_{t=2}^{S_p} \left\| \psi^{-1}(\mathbf{g}(t)) - \psi^{-1}(\mathbf{K}^{(t-1)}\mathbf{g}(1)) \right\|^2 \label{eq:pred} \\
\bz(t) & = \varphi(\bx(t), \bA(t)) \\
\bg(t) & = \phi(\bz(t)), \bg(t+1) = \bK\bg(t).
\end{align}
In \eqref{eq:lossre}, $\tilde \bx(t) =\mathbf{x}(t) - \hat{\mathbf{x}}(t)$ is the graph encoder-decoder's reconstruction error, and $\tilde \bz(t) =\mathbf{z}(t) - \hat{\mathbf{z}}(t)$ is the KAE encoder-decoder's reconstruction error, where $\hat{\mathbf{x}}(t)$ and $\hat{\mathbf{z}}(t)$ represent the ground truth values.
The term $S_p$ is a hyperparameter which decides the linearity length, defined during the training. 
Here,
$L_{\text{pred}}$ in \eqref{eq:pred} denotes the forward prediction loss, which assesses the accuracy of predicting future graph latent spaces and $L_{\text{rec}}$ in \eqref{eq:lossre} represents the reconstruction loss, which is utilized to evaluate the error of reconstructing the graph node features,  $\tilde{\bx}(t)$, and error of reconstructing the graph latent spaces,  $\tilde{\bz}(t)$.

\begin{table}[t]
\centering
\caption{Simulation Parameters for UAV Dynamics.}
\begin{tabular}{ll}
\toprule
Parameter & Value \\
\midrule
Area of operation & $5000 \times 5000 \ \text{m}^2$\\
Time step ($\Delta t$) & $0.1\ \text{s}$\\
Num. of UAVs (L) & 4 \\
Constant velocity ($\{v_1, v_2, v_3, v_4\}$) & $\{20, 20, 20, 20\}\ \text{m/s}$\\
Turning radius ($\{\phi_1, \phi_2, \phi_3, \phi_4\}$) & $\{0.25, 0.25, 0.25, 0.25\} \ \text{rad.}$ \\
Wind direction ($\theta_w$) & $10^{-8} \ \text{rad.}$\\
Wind speed ($v_w$) & $10^{-3} \  \text{m/s}$\\
Distance threshold ($\tilde{D}$) & $10^{4}\ \text{m}$\\
\bottomrule
\end{tabular}
\label{tab:UAVdynamics}
\end{table}

\begin{table}[t]
\centering
\caption{Simulation Parameters for LPD Problem.}
\begin{tabular}{ll}
\toprule
Parameter & Value \\
\midrule
Number of Ground Nodes ($N$) & 25\\
Maximum Transmit Power (\(P_{\text{max}}\)) & 0.1 W \\
Minimum Required Number of Communication Links (\(\tilde{C}\)) & 5 \\
Target SNR ($\tilde{\gamma}$) & 10 dB \\
Threshold for Received Power (\(\tilde{P}_{\text{det}}\)) & 0.5 \(\mu \text{W}\) \\
Path-loss constant ($\eta$) & $5$ \\
Path-loss constant for air-ground channel ($\eta^\prime$) & $2$ \\
Noise variance ($N_0$) & -174 dBm/Hz \\
\bottomrule
\label{tab:SP2}
\end{tabular}
\end{table}

\subsection{Transmit Power Optimization for LPD}

We aim to minimize the received power for the ground nodes in the terrestrial ad-hoc network at the predicted surveillance locations of UAVs. The optimization problem where we aim to minimize the maximum received power with the predicted locations of the $L$ UAVs is given by
\begin{align}
\textbf{(P2) :}  \quad \min_{P_n(t)} \max_{l}\ &  P_n(t) d_{l, n}^{-\eta}(t) \label{subeq:A}\\
\text{subject to} \quad
0 \leq P_n(t) &\leq P_{\text{max}}, \quad \forall n \in \mathcal{N} \label{subeq:pp} \\
 |C_n(t)| & \geq \tilde{C} \label{subeq:3}, \quad \forall n \in \mathcal{N} \\
P_{l, n}(t) & \le \tilde{P}_{\text{det}}, \quad \forall n \in \cN. \label{subeq:4}
\end{align}
In \eqref{subeq:A}, the transmit power of the ground nodes are to be optimized with the power constraint in \eqref{subeq:pp}. To avoid any isolated clusters, each ground node should have at least $\tilde{C}$ communication links, as seen in \eqref{subeq:3}. More importantly, in \eqref{subeq:4}, we enforce that the transmit power should be optimized such that the received power at the UAV is always less than a threshold, denoted by, $\tilde{P}_{\text{det}}$. This problem is NP-hard due to the constraints.
To reduce complexity, we assume that the same transmit power is used uniformly across all the nodes, meaning $P_n(t) = P(t), \forall n$, resulting in the simplified problem below:
\begin{align}
(\textbf{P2'}): \min_{P(t)} \max_{l} \ P(t) d_{l, n}^{-\eta}(t) \\
\text{subject to} \quad
\eqref{subeq:pp}, \eqref{subeq:3},\text{ and }\eqref{subeq:4}. \nonumber
\end{align}

\section{Numerical Results}\label{sec:SP}

In this section, we present numerical results for LPD with 25 ground nodes and 4 surveillance UAVs based on our proposed GKAE. Unless otherwise specified, simulation parameters follow Tables~\ref{tab:UAVdynamics} and ~\ref{tab:SP2} for UAV dynamics and LPD optimization, respectively.

\begin{figure}[t]
   \centering
   \begin{subfigure}{.49\columnwidth}  \includegraphics[width=\linewidth]{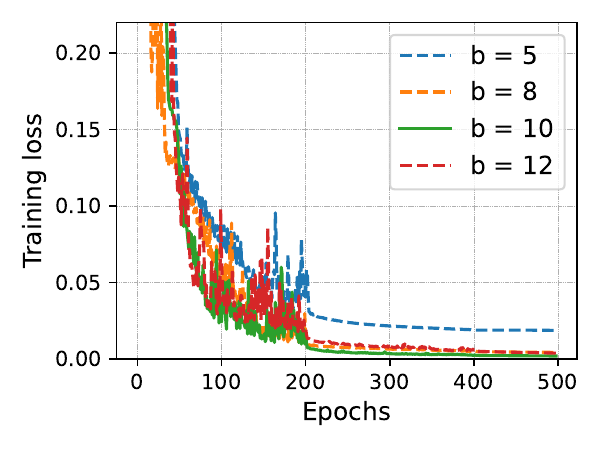}
      \caption{Train convergence}
      \label{fig:sub1}
   \end{subfigure}
   \hfill
   \begin{subfigure}{.49\columnwidth}
    \includegraphics[width=\linewidth]{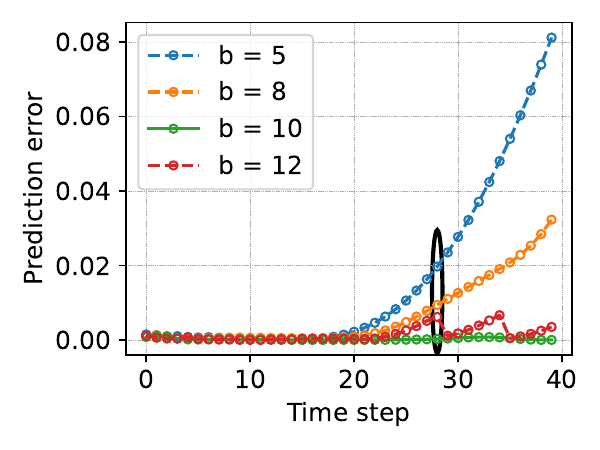}
      \caption{Prediction error}
      \label{fig:sub2}
   \end{subfigure}
   \caption{Training convergence and prediction errors of GKAE with respect to different Koopman matrix of dimension $b$.}
   \label{fig:combined}
\end{figure}

\begin{figure}[t]
    \centering
\includegraphics[width=.70\columnwidth]{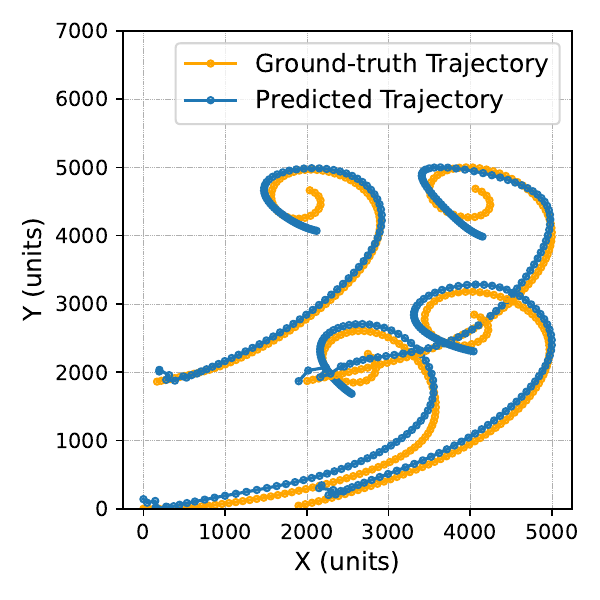}
    \caption{Predicted 4 UAV trajectories for $p = 80$ time steps.}
    \label{fig:Pred}
\end{figure}

\textbf{Trajectory Prediction}:
For predicting the trajectory of UAVs, we first train the GKAE architecture which consists of 3 SAGE layers and 6 fully-connected layers. The exponential linear unit (elu) and tangent hyperbolic (tanh) activation functions have been used after SAGE layers and fully-connected layers, respectively. The encoder-decoder architectures are connected through a linear fully-connected layer, i.e. a Koopman matrix. The GKAE is trained over 500 epochs using the Adam optimizer, given the loss function \eqref{eq:minloss} with $\beta_1 = \beta_2 = 1$ and $S_p = 20$. 
The GKAE performance depends significantly on the dimension of the Koopman matrix, which determines the number of eigenvalues in the Koopman invariant subspace. In Fig.~\ref{fig:sub1}, $M = b$ represents the chosen values for the dimension of the Koopman matrix. It is seen that the GKAE converges for various $b$ values but a very small value for $b$ decreases convergence speed. After training completes, Fig.~\ref{fig:sub2} displays the prediction error $\epsilon_{\text{pred}}$, measured using the mean-squared error averaged over a maximum prediction $p=40$ time steps, i.e., $\epsilon_{\text{pred}} = \sum_{t = 2}^p||\bx(t) - \varphi^{-1}(\psi^{-1}(\bK^{(t-1)}\bg(1)))||^2$. Here, the prediction input observation is fixed as $\bg(1))$ at the first time step. The results show that there $b=10$ achieves the lowest prediction error. Furthermore, with $b=10$, we extend the maximum prediction time steps to $p=80$, and obtain $\epsilon_{\text{pred}} = 0.0025$. The resulting predicted trajectories are close to the ground-truth values as visualized in Fig.~\ref{fig:Pred}.



\begin{figure*}[t]
   \centering
   \begin{subfigure}{0.31\linewidth}
      \includegraphics[width=\linewidth]{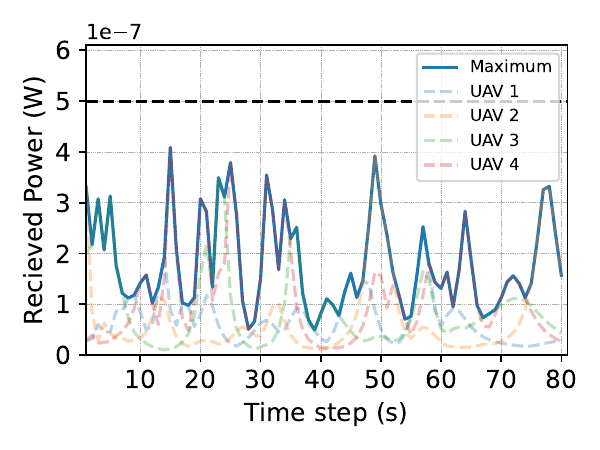}
      \caption{Received power at UAVs over varying time steps}
      \label{fig:Power}
   \end{subfigure}
   \hfill
   \begin{subfigure}{0.31\linewidth}
      \includegraphics[width=\linewidth]{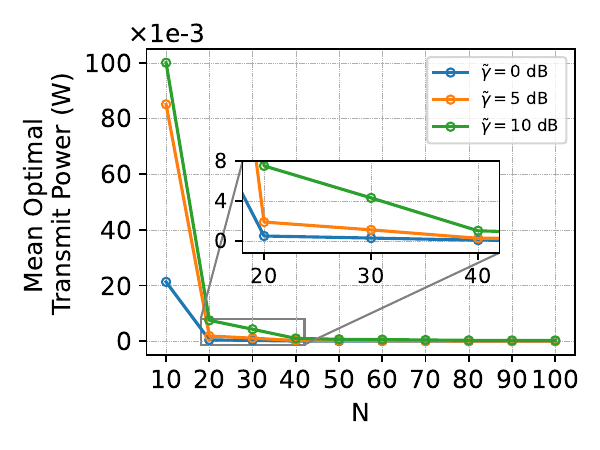}
      \caption{Mean transmit power over a varying number $N$ of ground nodes.}
      \label{fig:VaryingSNR}
   \end{subfigure}
   \hfill
   \begin{subfigure}{0.31\linewidth}
      \includegraphics[width=\linewidth]{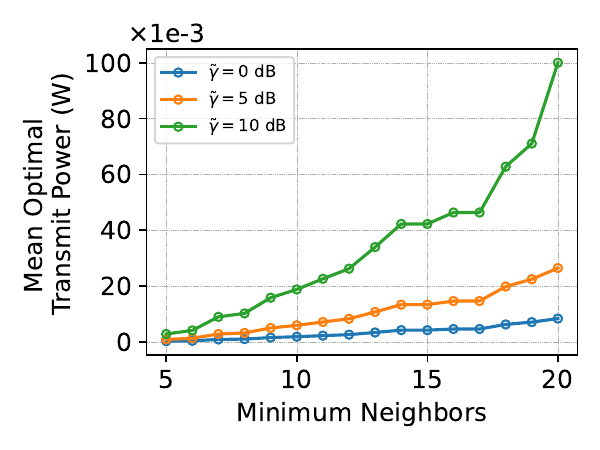}
      \caption{Mean transmit power over a varying minimum number of required links.}
      \label{fig:varyingD}
   \end{subfigure}
   \caption{Received power at UAVs, transmit power with respect to the number $N$ of ground nodes, and to the minimum number $\tilde{C}$ of required communication links.}
\end{figure*}

\textbf{Optimal Transmit Power}:
Fig.~\ref{fig:Power} visualizes the received power at UAVs over time when ground node transmit power is optimized as $P(t)^\ast$ by solving \textbf{P2'}. The result shows that the proposed solution ensures that a maximum received power remains below the UAV's detection threshold $\tilde{P_{\text{det}}}$ (dashed black line). This effectively ensures that UAVs conducting surveillance are unable to detect any unusual activity, as the received power is maintained at a very low level. To show the impact of the number $N$ of ground nodes, Fig.~\ref{fig:VaryingSNR} demonstrates the mean optimal transmit power $\mathbb{E}[P(t)^\ast]$ averaged over the maximum prediction time steps $p$. For different values of SNR threshold $\tilde{\gamma}$, the result consistently indicates that the mean optimal transmit power decreases rapidly with $N$ as the network density increases.

\begin{figure}
   \begin{subfigure}[b]{0.48\linewidth}
      \includegraphics[width=\linewidth,height=\linewidth]{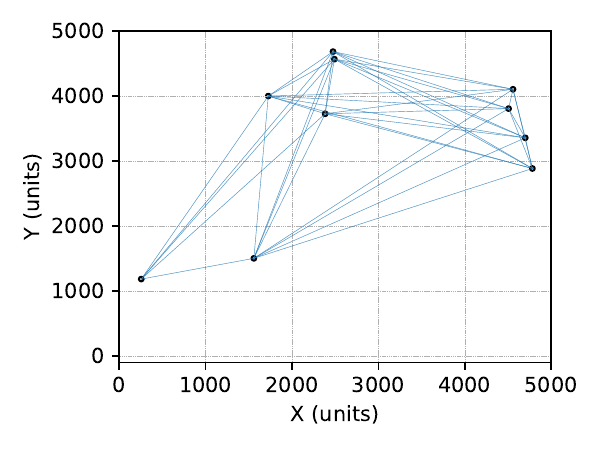} 
      \caption{$N = 10$}
      \label{fig:10}
   \end{subfigure}
   \hfill
   \begin{subfigure}[b]{0.48\linewidth}
      \includegraphics[width=\linewidth,height=\linewidth]{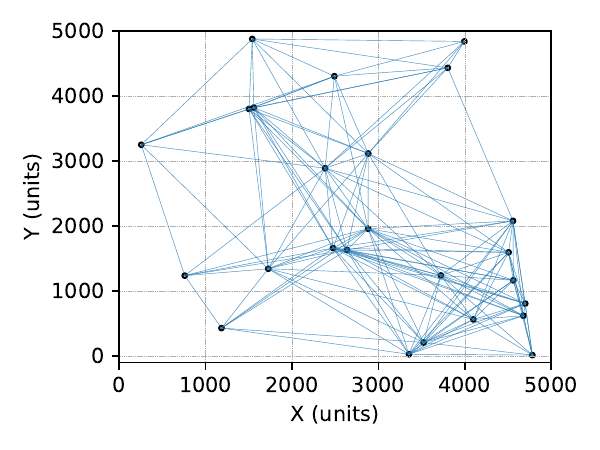} 
      \caption{$N = 25$}
      \label{fig:25}
   \end{subfigure}  
   \caption{LPD achieved network topology, given a minimum number of neighbors for each node with $\tilde{C} = 5$ and a target SNR of $\tilde{\gamma} = 10$ dB.}
   \label{fig:top}
\end{figure}

\textbf{Network Connectivity}:
Fig.~\ref{fig:varyingD} shows that the mean optimal transmit power increases not only with the SNR treshold $\tilde{\gamma}$ but also with the minimum number $\tilde{C}$ of required communication links. Note that our connectivity constraint imposes the minimum number of links per node, which may not always guarantee the network's full-connectivity without any isolated node clusters. To study this, we visualize the entire network topology in Fig.~\ref{fig:top}. With $\tilde{C}=5$, Fig.~\ref{fig:10} \text{and} Fig.~\ref{fig:25} show that the ground network become fully connected for both $N=10$ and $N=25$.




\section{Conclusions}\label{sec:Conc}

Our study has tackled the challenge of enabling LPD communication for terrestrial ad-hoc networks under UAV surveillance. To gain a full knowledge of the UAV mobility pattern, we aimed to predict the trajectory of the UAV surveillance based on a novel data-driven approach that integrates graph learning with Koopman theory. By leveraging GNN architecture, it was possible to learn the intricate spatial interactions in the UAV network and linearize the dynamics of multiple UAVs using the same architecture. Using these predicted locations, we conducted a case study for optimizing nodes' transmit power in a terrestrial ad-hoc network for minimizing detectability of RF signals. Extensive simulations have demonstrated accurate long-term predictions of fixed-wing UAV trajectories, which, in turn, hold promise for enabling low-latency LPD covert operations.
\bibliographystyle{ieeetr}
\bibliography{Koopman}

\end{document}